\documentclass[aps,pre,showpacs,twocolumn]{revtex4}
\usepackage{amssymb}
\usepackage{graphicx}
\usepackage{colordvi}
\usepackage{color}
\usepackage{dcolumn,amsmath,amsthm,amscd,amsfonts,amssymb,epsfig,graphics,graphicx,eucal}
\usepackage[colorinlistoftodos]{todonotes}

\begin{document}

\title{Intrinsic bistability and dual-core dark solitons and vortices  in exciton polariton condensates}
\author{A.V. Yulin$^{1}$, D.V. Skryabin$^2$, and A.V. Gorbach$^{2}$}
\affiliation{ $^1$ITMO University 197101, Kronverksky pr. 49, St. Petersburg, Russian Federation
\\
$^2$Department of Physics, University of Bath, Claverton Down, Bath, BA2 7AY, UK}

\date{\today}

\begin{abstract}
We investigate a new kind of dark solitons and vortices that can exist in the exciton-polariton condensates. These structures have discontinuity in the excitonic part of the polaritonic field and exist due to an intrinsic multiplicity of the solutions for the exciton density in the given optical field. Reported solutions are characterized by two very distinct localization scales, and hence are coined as dual-core dark solitons and vortices. 
\end{abstract}
\maketitle

\input{epsf.tex}
\epsfverbosetrue

\section{Introduction}
\label{intro}

Photons in optical microcavities can strongly couple to excitons and form new quasi-particles:  exciton-polaritons or simply polaritons \cite{model_Kavok,model_book}. These unique half-light half-matter states exhibit rich physical properties and attract great interest from condensed matter and optics communities. Inheriting a light effective mass from its photon component, microcavity polaritons demonstrate high temperature Bose-Einstein condensation  \cite{kaspr, balili} and superfluidity \cite{hm_sol_Carusotto, amo_sf}. At the same time, due to their excitonic component, microcavity polaritons interact much stronger than pure photons in typical photonic setups, and thus represent a competitive and promising platform for ultrafast signal processing applications. Indeed, a range of fundamental nonlinear effects has been investigated with microcavity polaritons, including low threshold bistability \cite{bistab, bistab2}, polarization multi-stability and switching \cite{bistab_pol1, bistab_pol2, bistab_pol3}, parametric scattering and pattern formation \cite{savvidis, hm_sol_Wouters, vortex_lattice, hm_sol_Saito, hm_sol_Luk}, excitation of dark- \cite{darkS_Amo, soliton_Yulin, dark_Werner, SmirnovKivshar_PRB, darkS_Pinsker} and brigth solitons \cite{soliton_EgorovPRL, solit_Egorov2010, solit_Egorov2012, soliton_EgorovBook, soliton_Sich, soliton_EgorovPRB, soliton_SichPRL}.

The existence of dark solitons is supported by the combination of the positive effective mass of polaritons with low momenta in the cavity plane, and the repulsive nonlinear interaction of polaritons. The two factors jointly ensure stability of a high amplitude background, which forms the tails of a dark soliton \cite{soliton_Yulin}. In planar systems, such as polariton microcavities, these conditions also favor formation of vortices, and the two types of solutions- dark solitons and vortices, often coexist \cite{soliton_OstrovskayaPRA}. In particular, vortices can emerge as the result of a transverse instability development of a dark soliton \cite{SmirnovKivshar_PRB}.
Spontaneous parametric scattering can also lead to formation of various types of rotating and stationary vortex-antivortex lattice arrangements in microcavities \cite{vortex_lattice, vortex_Keeling, vortex_Marchetti, vortex_Borgh, vortex_Manni}. Excitation of vortices can be triggered by perturbations, such as an external pulse or polariton condensate scattering by an obstacle \cite{vortex_Liew}. The latter mechanism underlines similarities of polariton dynamics in microcavities with hydrodynamics of super-fluids \cite{super_Tilley}. The analogy has been further developed in the recent theoretical and experimental studies of hydrodynamical properties of vortices and dark solitons in polariton condensates \cite{vortex_Pigeon, vortex_Grosso, solit_Grosso1, vortex_Saito}. Despite the above analogies with hydrodynamics and optics, polariton solitons and vortices have an important distinctive feature: they are inherently two-component states, and the effective masses of the photonic and excitonic components differ by many orders of magnitude. The fact that there are two components is somewhat obscured, when polaritons are described by a single order parameter function, but in this work we demonstrate that there are easily accessible regimes and solutions, where this approach breaks down.

In this paper, we demonstrate that the dual-component nature of micro-cavity polaritons leads to the existence of the {\em novel type} of dark solitons and vortices which have no analogues in hydrodynamics, nonlinear optics and dynamics of atomic Bose-Einstein condensates. Such solutions emerge as the result of the specific type of bistability of the system -  so called {\em intrinsic bistability}. This phenomenon has been known for sometime, though very rarely investigated, and it refers to the situation when material polarisation  is bistable in the presence of a constant applied field \cite{intrinsic}. In our system, the intrinsic bistability manifests in multiple solutions of the excitonic wave function for a given amplitude of the intracavity electric field. We show that the intrinsic bistability provides the new and unique mechanism for localization of nonlinear excitations. We coin such solutions {\em dual-core} vortices and solitons. 
Recently it was noted that polariton vortices have two distinctive characteristic lengths of localization (healing lengths) in the photonic and excitonic components \cite{vortex_Voronova}. However, vortices and dark solitons reported in that work 
can be described within the framework of the polaritonic order parameter equation of the Gross-Pitaevskii type, and exist outside the parameter range where the intrinsic bistability and our solutions are possible.
We also note here the analogy with Abrikosov vortices in superconductors of the second type placed in  magnetic field \cite{super_Tilley, super_Tinkh}. Abrikosov vortices  have two very different scales: the so-called London length, which is the relatively large characteristic scale of the localization of magnetic field, and the coherence length, which is the small characteristic length of the variation of the superconducting electrons density.

\section{Model equations}
\label{sec:model}

We adopt the well established dimensionless mean-field model describing dynamics of the exciton polariton condensate in an optical cavity \cite{model_Kavok,model_book} in terms of coupled complex amplitudes $E$ and $\psi$ of the photon and exciton fields:
\begin{eqnarray}
\label{main_equations}
\partial_t E -i \nabla^2_{\perp} E+(\gamma_1 -i\delta)E= i\psi +E_p \cdot e^{-i q_p t}\\
\partial_t \psi -i\sigma \nabla^2 \psi + (\gamma_2 +i\delta+i|\psi|^2)\psi= iE,
\label{main_equations_}
\end{eqnarray}
where time is measured in the units of the inverse Rabi frequency $T=1/\omega_R$, scaling of spatial coordinates $L$ is determined by the effective cavity photon mass $m_c$: $L=\sqrt{\hbar/(2m_c\omega_R)}$, $\gamma_1$ and $\gamma_2$ are the attenuation coefficients for the photon and exciton fields, respectively, $2\delta$ is the detuning between the exciton resonance and the cavity resonance, $\sigma$ is the coefficient describing the diffraction of the exciton field (relative to the diffraction of photons): $\sigma\sim m_{c}/m_{ex}\ll 1$, $m_{ex}$ is the exciton mass,
$E_p$ and $q_p$ are the amplitude and the detuning of the pump from the center of the gap between lower- and upper polaritonic branches.
The coefficient of nonlinear interaction in the excitonic field is set to unity by the appropriate scaling of field amplitudes,
see more details in Ref. \cite{soliton_Sich}.

For zero pump, $E_p=0$, and neglecting dissipation and nonlinearity, the spectrum of low-amplitude zero momentum polaritons $E,\psi\sim e^{-i\omega t}$ is given by
\begin{equation}
\omega=\pm\sqrt{1+\delta^2}\;.
\end{equation}
This defines the gap between the lower and upper polariton branches, see Fig.~\ref{fig_1_1}. Note, that Voronova et al assumed $\delta=0$ in Ref. \cite{vortex_Voronova}, which makes impossible existence of the dual-core solitons discussed below.

\section{Intrinsic bistability and stationary solutions}

In this section we consider stationary solutions, $E=A e^{-iqt}, \psi=\Psi e^{-iqt}$, in the conservative limit of negligible dissipation and zero pump: $\gamma_1=\gamma_2=E_p=0$. Amplitudes $A$ and $\Psi$ solve the following set of equations:
\begin{eqnarray}
\label{2d_stat_equations1}
\left(\partial_{x}^2+\partial_{y}^2\right)A+(q+\delta) A+\Psi=0 \;, \\
\sigma \partial_{x}^2 \Psi +(q-\delta-\Psi^2)\Psi +A=0\;.
\label{2d_stat_equations2}
\end{eqnarray}

Neglecting the diffraction term in the excitonic field, $\sigma=0$, Eq.~(\ref{2d_stat_equations2}) becomes the cubic algebraic equation:
\begin{equation}
(q-\delta)\Psi-\Psi^3+A=0\;.
\label{eq:algebraic}
\end{equation}
Generally, for a given amplitude $A$ of the photonic component, it admits either one or three real solutions: $\Psi=\Psi_m(A)$, $m=1,2,3$. 

We define  coexistence of the {\em three distinct states of excitons for the same amplitude of the photon field} as the {\em intrinsic bistability} of the system. We emphasize, that the intrinsic instability refers to the excitonic field only, and it is different from the conventional bistability of homogeneous solutions of the coupled system of Eqs. (\ref{main_equations}) and (\ref{main_equations_}).

\begin{figure}[h]
\vspace{0.5cm}
\epsfig{file=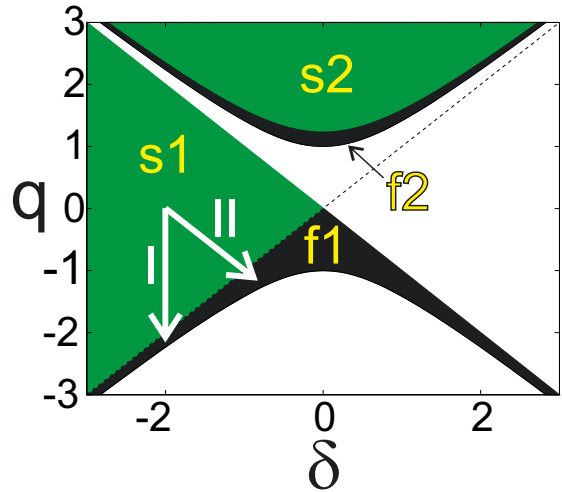,angle=0,width=0.9\columnwidth}
\caption{(Color online) Domains of existence of the non-trivial spatially homogeneous solution (shaded areas $s1$, $s2$, $f1$, $f2$), and intrinsic bistability domains (green areas $s1$ and $s2$) in the plane of parameters $(q,\delta)$.}
\label{fig_1_1}
\end{figure}

The non-trivial spatially homogeneous solution of Eqs.~(\ref{2d_stat_equations1}, \ref{2d_stat_equations2}) with $A(x,y)=A_h=const.$, $\Psi(x,y)=\Psi_h=const.$ is given by:
\begin{equation}
\Psi^2_h=\frac{q^2-\delta^2-1}{q+\delta}\;,\qquad A_h^2=\frac{\Psi_h^2}{(q+\delta)^2}.
\label{eq:psi_sq_solution}
\end{equation}
The domains of existence of this solution in the plane of parameters $(q,\delta)$ are indicated in Fig.~\ref{fig_1_1} as four shaded areas $s1,s2$ and $f1,f2$.

Analysis of Eq. (\ref{eq:algebraic}) shows that it admits three real solutions when $\Psi^2<(q-\delta)/3$. Combining this result with the solution for $\Psi^2$ in Eq.~(\ref{eq:psi_sq_solution}), we obtain the condition for the intrinsic bistability:
\begin{equation}
2(q-\delta)>\frac{3}{q+\delta}>0\;.
\label{eq:intrinsic_bi_condition}
\end{equation}
In Fig.~\ref{fig_1_1} the domains of the intrinsic bistability  are marked as $s1$ and $s2$ (green), 
while domains $f1$ and $f2$ (dark grey) correspond its absence. Homogeneous solution Eq. (\ref{eq:psi_sq_solution}) exists
($\Psi^2_h>0$) across all these domains.
Linear stability analysis shows that the homogeneous solution is dynamically unstable in $f2$ and $s2$ domains. We remark that it can be stabilized by including sufficiently high losses and pump, however such a strongly dissipative case is out of the scope of this paper and will be discussed elsewhere.
On the contrary, the homogeneous solution is linearly stable in $s1$ and $f1$ domains.  Hence it can serve as the background for possible dark solitons and vortices.

To describe localized solutions, we now consider spatially nonuniform amplitudes $A$ and $\Psi$. For $\sigma=0$, with the account of three possible solutions, $\Psi_m$, of the algebraic equation (\ref{eq:algebraic}), Eqs.~(\ref{2d_stat_equations1}, \ref{2d_stat_equations2}) can be written as the  equation for the photonic component $A$:
\begin{eqnarray}
\label{2d_stat_equiv}
(\partial_{x}^2+\partial_{y}^2) A+(q+\delta) A+\Psi_m(A)=0\;, \;\; m=1,2,3\;.
\end{eqnarray}

Let us first discuss the 1D case in which the fields are homogeneous along $y$-coordinate. Then, Eq.~(\ref{2d_stat_equiv}) is equivalent to the equation describing dynamics of a particle in a potential: 
\begin{equation}
\partial^2_x A= -\frac{\partial F_m}{\partial A}\;,
\end{equation}
where coordinate $x$ plays the role of the effective "time", and the potential is given by:
\begin{equation}
F_m(A)=\frac{3\Psi_m^4(A)}{4}-\frac{(q-\delta) \Psi_m^2(A)}{2}+\frac{(q+\delta) A^2}{2}\;.
\label{potent}
\end{equation}
In the derivation of Eq.~(\ref{potent}) we used $\partial/\partial A=\partial/\partial \Psi_m \cdot \left(\partial A/\partial \Psi_m\right)^{-1}$, where $\partial A/\partial \Psi_m$ is obtained by direct differentiation of Eq.~(\ref{eq:algebraic}).

\begin{figure}[h]
\vspace{0.5cm}
\epsfig{file=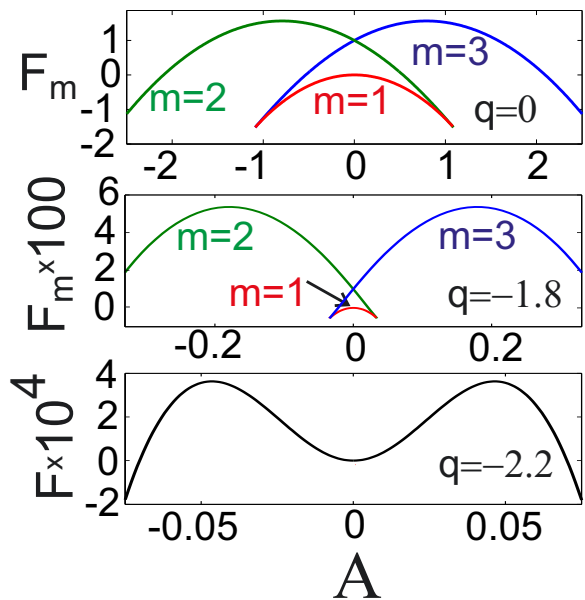,angle=0,width=0.9\columnwidth}
\caption{(Color online) The effective potential $F_m$, Eq.~(\ref{potent}), for $q=0$, $q=-1.8$ and $q=-2.2$. The detuning is fixed to $\delta=-2$.}
\label{fig_1_1_bis}
\end{figure}

In Fig.~\ref{fig_1_1_bis} the potential $F_m(A)$ is plotted using different roots $\Psi_m(A)$ for a set of values of $(q-\delta)$. As follows from Eq.~(\ref{eq:algebraic}), the potential is multi-valued within the range of the photon field amplitude $|A|<2\left[(q-\delta)/3\right]^{3/2}$. As $(q-\delta)$ tends to zero, the overlap area shrinks, and the potential becomes a single-valued function, see the bottom panel in Fig.~\ref{fig_1_1_bis}.

\begin{figure}[h]
\vspace{0.5cm}
\epsfig{file=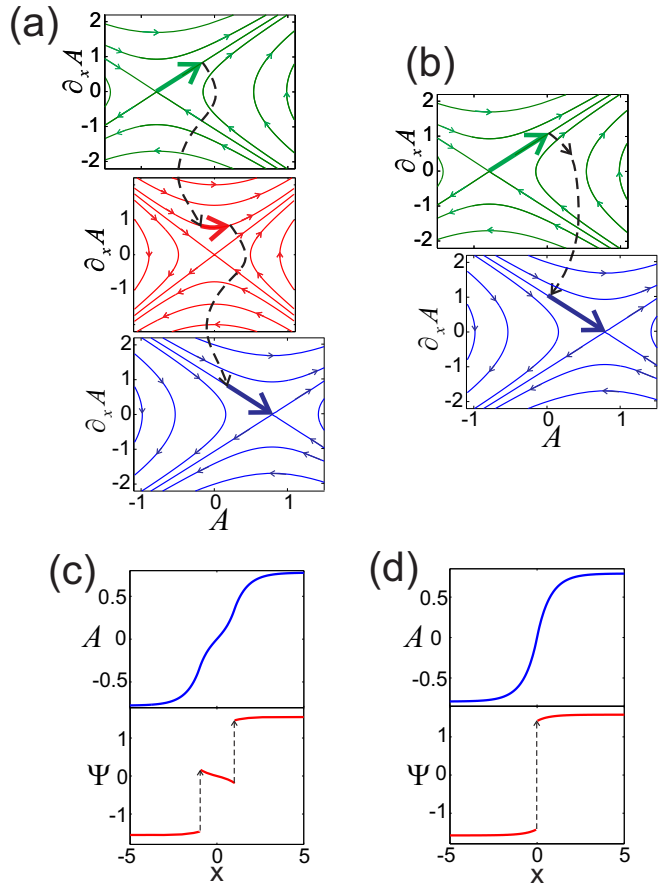,angle=0,width=\columnwidth}
\caption{(Color online) The phase planes for $q=0$, $\delta=-2$. Panels (a) and (c) illustrate the solution with two symmetric discontinuities at $x=\pm 1$. A special case with the direct switch from the upper to the bottom phase plane is shown in panels (b) and (d).}
 \label{fig_2}
\end{figure}

The corresponding phase planes for $q=0$ and $\delta=-2$ [i.e. $(q-\delta)=2$, 
see the top panel in Fig.~\ref{fig_1_1_bis}] are plotted in Fig.~\ref{fig_2}(a). Each phase plane, corresponding to a separate root $\Psi_m$, has one saddle fixed point. Crucially, for $\sigma=0$ the equation for $\Psi$ is algebraic, and hence the $\Psi$ {\em field can be discontinuous}. At the same time, continuity of the $A$ field and its derivative $\partial_x A$ is still preserved by virtue of the differential equation (\ref{2d_stat_equiv}). In other words, within the range of amplitudes $|A|<2\left[(q-\delta)/3\right]^{3/2}$ where the three phase planes overlap, a phase trajectory can switch from one $\Psi_m$ to another, provided that continuity of $A$ and $\partial_x A$ is preserved. This allows us to construct various localized solutions, for which the phase trajectory connects two saddle points from different phase planes. 

An example of one such phase trajectory is illustrated in Fig.~\ref{fig_2}(a). At $x\to -\infty$ the trajectory starts in the saddle point of the phase plane corresponding to the root $\Psi_2(A)$, see top panel in Fig~\ref{fig_2}(a). Within the allowed range of $A$, the trajectory switches to  the phase plane corresponding to the root $\Psi_1(A)$ and  then jumps to the $\Psi_3$ one, where it terminates at the saddle point, see the bottom panel of  Fig.~\ref{fig_2}(a). The corresponding solution is shown in Fig.~\ref{fig_2}(c) and it represents a dark soliton. The field $A$ is continuous, but the field $\Psi$ has discontinuities at two points, where the switches to different phase portraits occur.
The phase trajectory can also switch directly between roots $\Psi_2(A)$ and $\Psi_3(A)$, as illustrated in Fig.~\ref{fig_2}(b). The corresponding soliton solution is plotted in Fig.~\ref{fig_2}(d). Comparing $A$ and $\Psi$ field profiles in Fig.~\ref{fig_2}(d), it is obvious that the soliton core in the $A$ field is much larger than in the $\Psi$ field, thereby forming a dual core dark soliton.
We note, that the switching between the two roots occurs at a particular value of the field amplitude $A_0$, for which the continuity of $\partial_x A$ is preserved. For fixed values of parameters $q$ and $\delta$, this condition uniquely defines the structure of the soliton, in particular its localization lengths in each of the field components, and the value of $\Psi$ field at the discontinuity. 

\begin{figure}[h]
\vspace{0.5cm}
\epsfig{file=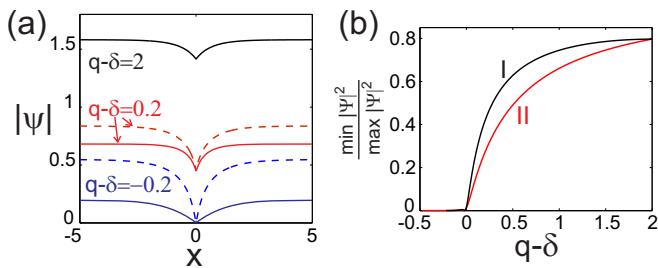,angle=0,width=\columnwidth}
\caption{(Color online) Transition from dual-core to conventional dark solitons.
In panel (a) profiles of dark solitons are plotted for different values of $(q-\delta)$. Solid curves correspond to the case of fixed $\delta=-2$, dashed curves to the case of fixed $q+\delta=-2$.
Panel (b) shows the dependencies of the inverse contrast of $|\Psi|$ in a dark soliton on $(q-\delta)$. The curve marked as $I$ is calculated for fixed $\delta=-2$, the curves marked as $II$ is calculated for $q+\delta=-2$, see the corresponding white arrows in Fig.~\ref{fig_1_1}.}
\label{fig_1_2}
\end{figure}

It is instructive to consider how the above {\em dual core dark solitons} transform into conventional dark solitons, which exist for $(q-\delta)<0$, see domain $f2$ in Fig.~\ref{fig_1_1}. We note that $\Psi_2(A)=-\Psi_3(A)$, and therefore the profile of $|\Psi|$ for the dual core soliton in Fig.~\ref{fig_2}(d) is continuous, see Fig.~\ref{fig_1_2}(a). In the soliton core, the field $|\Psi|$ reaches its minimum, however the minimum is always larger than zero and corresponds to $|\Psi_2(A_0)|=|\Psi_3(A_0)|$. Decreasing $(q-\delta)$ and approaching the boundary between the $s1$ and $f1$ domains, the range of  $A$, where the overlap between the three phase planes occurs, gradually shrinks. As a result, the soliton background drops, as well as the minimum of $|\Psi|$ field in the soliton core. To characterize the discontinuity in the soliton core, in Fig.~\ref{fig_1_2}(b) we plot the ratio between the minimal and the background values of $\Psi$ as functions of $(q-\delta)$. Curves $I$ and $II$ correspond to two different paths in the parameter plane $(q,\delta)$, as indicated with white arrows in Fig.~\ref{fig_1_1}. Approaching the boundary between $s1$ and $f1$ domains, this ratio gradually reduces, but remains non-zero everywhere inside $s1$ domain, thus confirming that the soliton has a discontinuity of the  $\Psi$ field in its core.

Crossing the boundary between $s1$ and $f1$ domains, the relationship between $\Psi$ and $A$ becomes unique (no intrinsic bistability). Inside $f1$ domain, the potential $F$  and the corresponding phase plane   now have three fixed points: the centre point at $A=0$ and two saddles at $A=\pm |A_h|$. In this domain, usual dark solitons \cite{vortex_Voronova}, corresponding to heteroclinic orbits, exist. The profile of $\Psi$ field becomes continuous and crosses zero in the soliton core. Therefore, the minimum value of $|\Psi|$ becomes equal to zero for such solitons, and the ratio between the minimum and background values is always zero, see Fig.~\ref{fig_1_2}.

\begin{figure}[h]
\vspace{0.5cm}
\epsfig{file=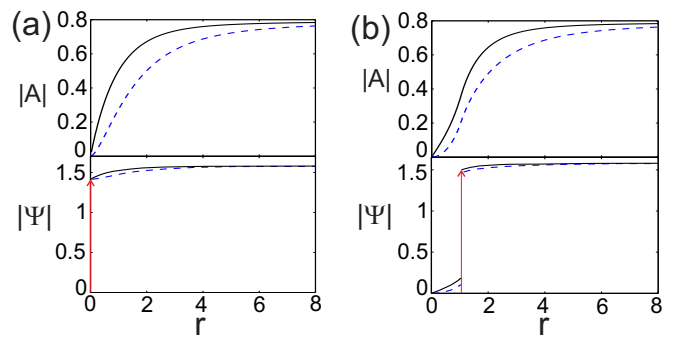,angle=0,width=\columnwidth}
\caption{(Color online) Radial profiles of dual-core vortex solitons with $m=1$ (solid curves) and $m=2$ (dashed curves). In panel (a) vortex solitons with the discontinuity in $\Psi$ field in the centre ($r=0$) are shown. The vortex solitons in panel (b) have discontinuity along the ring ($r=1$). The parameters are $\sigma=0$, $q=0$, $\delta=-2$.}
\label{2Dvortex}
\end{figure}

The described above dual-core dark solitons with discontinuities can be generalized to two dimensional vortices. Introducing polar coordinates $(r,\theta)$, and looking for radially symmetric solutions in the form $A=f(r)\exp(im\theta)$, $\Psi=p(r)\exp(im\theta)$, where $m$ is the topological charge of the vortex, it is easy to write the analogue of Eq.~(\ref{2d_stat_equiv}) for the functions $f(r)$ and $p(r)$. Numerical solutions of this equation corresponding to the dual-core vortices ($m=1$) localized at discontinuities in $\Psi$ field are shown in Fig.~\ref{2Dvortex}.  

\section{Stability analysis and dynamical evolution}

\begin{figure}[h]
\vspace{0.5cm}
\epsfig{file=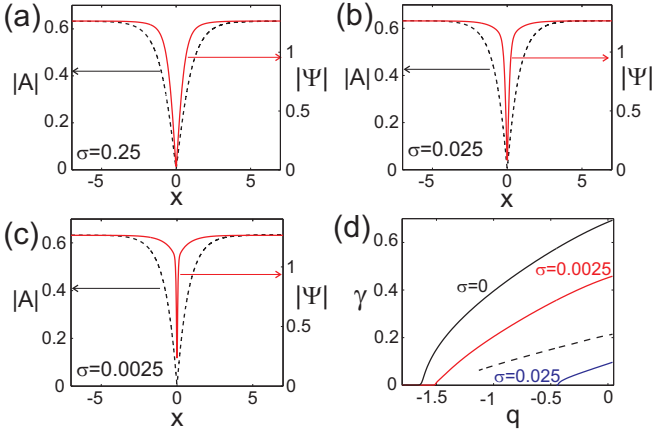,angle=0,width=\columnwidth}
\caption{(Color online) Dual-core dark solitons with non-zero diffraction in the excitonic component, $\sigma> 0$. Panels (a), (b) and (c) show the profiles of $A$ (dashed curves) and $\Psi$ (solid curves) fields for $\sigma=0.25$, $\sigma=0.025$ and $\sigma=0.0025$, respectively. Other parameters are: $\delta=1.55$, $q=-0.45$. 
In panel (d) the instability growth rates are plotted as functions of $q$ for the soliton with one discontinuity (solid curves) and $\sigma=0.25,0.025,0.0025$, and a soliton with two discontinuities separated by a distance $\Delta x = 1$, $\sigma=0$ (dashed curve).
}
\label{fig_4}
\end{figure}

Structural stability of the dark soliton solutions with discontinuities
on addition of the kinetic energy (dispersion) term in the exciton equation ($\sigma\ne 0$) is an important problem, which we address below. To investigate this issue we solved the 1D version of Eqs. (\ref{2d_stat_equations1}) and (\ref{2d_stat_equations2}) numerically for small $\sigma$. The results are shown in Fig.~\ref{fig_4}.  One can see that, as expected, the finite exciton dispersion removes the discontinuity in the $\Psi$ field, while the overall structure of solution is retained.
Thus dark solitons with discontinuities are structurally stable and transform into dual-core solitons for finite $\sigma$. To observe dark solitons and vortices in real physical experiments, the solutions must also be dynamically stable. We have studied the stability of the solitons by solving the corresponding spectral problem and by direct numerical simulation of Eqs. (\ref{main_equations})-(\ref{main_equations_}) with the noise added to the soliton profiles. The dependencies of the instability growth rates on the soliton parameter $q$ are shown in panel (d) of Fig.~\ref{fig_4} for different values of $\sigma$.

The linear stability analysis of conventional dark solitons reveals that they become unstable with respect to drift instability when  $q$ is increasing 
and the same instability is inherited by the dark solitons with discontinuities. Increasing 
dispersion in the exciton model tends to suppress   this instability, see Fig.~\ref{fig_4}(d).

Despite the instability of the dark soliton and vortices with very small diffraction in the exciton field, we demonstrate that similar structures can be experimentally observed in the decaying condensate. In the experiments polariton systems are dissipative having life time of order of $10$ ps. We performed numerical simulations of the 2D polariton system excited by a vortex laser beam and observed dual-core vortices in the decaying condensate. In this case the instability is suppressed by the dissipation and the observation time is limited by the lifetime of polaritons.

\begin{figure}[h]
\vspace{0.5cm}
\epsfig{file=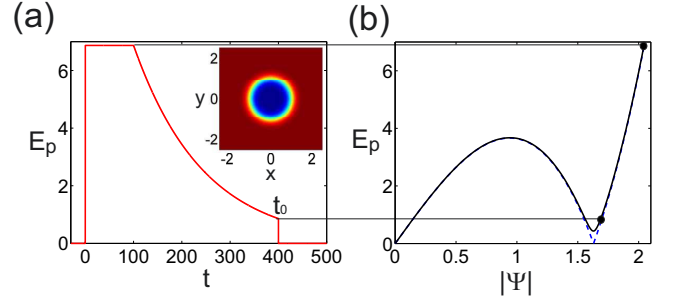,angle=0,width=\columnwidth}
\caption{(Color online) 
The variation of the pump amplitude $E_{p0}(t)$ in the numerical experiment is shown in panel (a). The parameters are $q_{p}=0$, $\delta=-2$. The inset shows the spatial distribution of the intensity of the pump. The bifurcation diagram for the spatially uniform backgrounds is shown in panel (b).}
\label{vortex decay2_1}
\end{figure}

To study formation of vortices in the decaying condensate we performed a numerical experiment exciting the system by an optical vortex beam with the topological charge one:
\begin{eqnarray}
E_p&=&E_{p0}(t) \times \nonumber \\
&&\times \left[ 2+\tanh\left(\frac{r-r_0}{w_p}\right)- 
\tanh\left(\frac{r+r_0}{w_p}\right) \right]\frac{x+iy}{2r}\;, \nonumber
\end{eqnarray}
where $r=\sqrt{x^2+y^2}$, for our numerical simulations we chose $r_0=1$, $w_p=0.25$. 
The corresponding spatial profile of the pump intensity is shown in the inset of Fig.~\ref{vortex decay2_1}.

\begin{figure}[h]
\vspace{0.5cm}
\epsfig{file=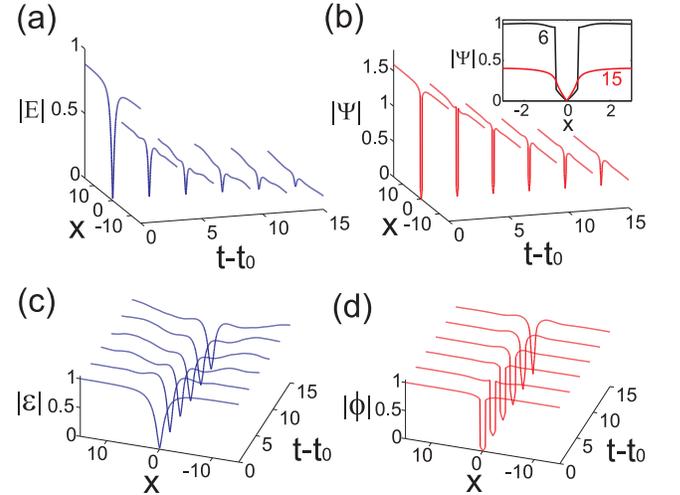,angle=0,width=\columnwidth}
\caption{(Color online)
Fields evolution in the numerical experiment: actual fields $E$ and $\Psi$, panels (a) and (b), and re-normalized fields $\varepsilon=|E|/\max(|E|$ and 
$\Phi=|\Psi|/\max(|\Psi|)$, panels (c) and (d). Parameter values are: $\sigma=0$, $q_{p}=0$, $\delta=-2$. The inset in panel (b) shows profiles of $|\Psi|$ field at $t-t_0=6$ and $t-t_0=15$, the curves are marked correspondingly.}
\label{vortex decay2_2}
\end{figure}

The amplitude $E_{p0}$ of the pump was adiabatically varied in time, as shown in panel (a) of Fig.~\ref{vortex decay2_1}. We switched the pump on at $t=0$ and choose the initial amplitude large enough to bring the vortex background to the high amplitude state, see panel (b) of Fig.~\ref{vortex decay2_1} showing the bifurcation diagram for spatially uniform states. Then we waited to obtain a stationary state, and at $t=100$ we started to decrease adiabatically the pump until at $t_0=400$ it reached the value $E_{p0}=0.43$ close to the lower folding point of bifurcation diagram for the spatially uniform solutions. 

Then we switched the pump off, which has led to the excitation of waves that affected mainly the $E$ filed. Due to dissipation the fields started to decay, see panels (a) and (b) of Fig.~\ref{vortex decay2_2}. 
In panels (c) and (d) of Fig.~\ref{vortex decay2_2} the corresponding re-normalized fields are plotted. The structure of the vortex is clearly preserved as fields decay.

\begin{figure}[h]
\vspace{0.5cm}
\epsfig{file=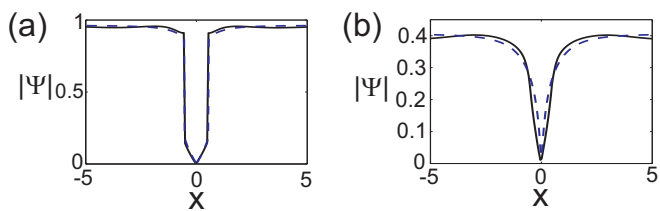,angle=0,width=\columnwidth}
\caption{(Color online) Field profiles in the numerical experiment for the times $t=6$ (a) and $t=15$ (b). Dashed curves show the stationary vortex solution for the conservative problem.}
\label{vortex decay_cmp}
\end{figure}

One can notice that initially the core in $\Psi$ field is much narrower compared to the one in $E$ field, as it should be in a dual-core vortex. With the decay of the background the both cores of the vortex broaden, but  the $\Psi$-core remains much  narrower, see the normalized field distributions in panels (c) and (d) of Fig.~\ref{vortex decay2_2}. The onset of a drastic broadening of the $\Psi$-core happens at the moment when adiabatic decrease of the vortex $q$-parameter transforms the dual core vortex into a usual one. 
In Fig.~\ref{vortex decay_cmp} the structure of a vortex in the decaying condensate is compared against the numerically exact vortex solution, both are clearly very similar.

\section{Conclusion}
We have demonstrated that intrinsic bistability of excitons leads to the existence of a new type of dark solitons and vortices. Without dispersion of the exciton field these structures have discontinuities in the exciton field. Arbitrary dispersion in the excitonic field removes these discontinuities and results in dark solitons and vortices, which have very different core sizes in the optical and excitonic fields. These double core vortices and solitons exist in the different parameter range relative to the previously studied polariton vortices, which can be well described by the single polaritonic amplitude (an order parameter) and have comparable core sizes in the excitonic and optical components \cite{vortex_Voronova}.
Formation of the dual core vortices in the decaying condensate under the realistic excitation conditions takes place despite possible presence of the dynamical instabilities, since the characteristic time of the instability development is larger than the life time of the polariton condensate.

\section*{Acknowledgements}
The work of AVY was financially supported by the Government of the Russian Federation (Grant 074-U01) through ITMO Early Career Fellowship scheme. All authors acknowledge financial support from the EU network project LIMACONA (Project No: 612600). DVS and AVG acknowledge funding through Leverhulme Trust Research Project Grant RPG-2012-481.


\begin{thebibliography}{99}


\bibitem{model_Kavok} A. V. Kavokin, J.J. Baumberg, G. Malpuech, and F.P. Laussy, {\it Microcavities} (Oxford University Press, Oxford,
2007).

\bibitem{model_book} {\it Exciton Polaritons in Microcavities }, edited by D. Sanvitto and
V. Timofeev, Springer Series in Solid-State Sciences Vol. 172(Springer, Heidelberg, 2012).

\bibitem{kaspr} J. Kasprzak {\em et al.}, Nature Physics {\bf 443}, 409 (2006).

\bibitem{balili} R. Balili, V. Hartwell, D. Snoke, L. Pfeiffer, and K. West, Science {\bf 316}, 1007 (2007).

\bibitem{hm_sol_Carusotto} I. Carusotto and C. Ciuti, Phys. Rev. Lett. {\bf 93}, 166401 (2004).

\bibitem{amo_sf} A. Amo {et al.}, Nature Phys. {\bf 5}, 805 (2009).


\bibitem{bistab}
A. Baas, J.P. Karr, H. Eleuch, and E. Giacobino, 
Phys. Rev. A {\bf 69}, 023809 (2004).

\bibitem{bistab2}
N. A. Gippius, S. G. Tikhodeev, V. D. Kulakovskii, D. N. Krizhanovskii, and
A. I. Tartakovskii,  Europhys. Lett. {\bf 67},
997 (2004).

\bibitem{bistab_pol1}
D. Sarkar {\em et al.}, Phys. Rev. Lett. {\bf 105}, 216402 (2010).

\bibitem{bistab_pol2}
A. Amo {\em et al.}, Nature Photon. {\bf 4}, 361 (2010).

\bibitem{bistab_pol3}
T. K. Paraiso, M. Wouters, Y. Leger, F. Morier-Genoud, and B. Deveaud-Pledran, 
Nature Mater. {\bf 9}, 655 (2010).

\bibitem{savvidis}
P. G. Savvidis {\em et al.} Phys. Rev. Lett. {\bf 84}, 1547 (2000).

\bibitem{hm_sol_Wouters} M. Wouters and I. Carusotto, Phys. Rev. B {\bf 75}, 075332 (2007).


\bibitem{vortex_lattice}
A. V. Gorbach, R. Hartley, and D. V. Skryabin, Phys. Rev. Lett. {\bf 104},
213903 (2010).

\bibitem{hm_sol_Saito} H. Saito, T. Aioi, and T. Kadokura, Phys. Rev. Lett. {\bf 110}, 026401 (2013).

\bibitem{hm_sol_Luk} M.H. Luk, Y.C. Tse, N.H. Kwong, P.T. Leung, P. Lewandowski, R. Binder, and S. Schumacher,
Phys. Rev. B {\bf 87}, 205307 (2013)




\bibitem{darkS_Amo} A. Amo, S. Pigeon, D. Sanvitto, V.G. Sala, R. Hivet, I. Carusotto, F. Pisanello, G. Lemenarger, R. Houdre, E. Giacobino, C. Ciuti, and A. Bramati, Science {\bf 332}, 1167 (2011).

\bibitem{soliton_Yulin}
A. V. Yulin,  O. A. Egorov, F. Lederer, and D. V. Skryabin, Phys. Rev. A \textbf{78}, 061801(R) (2008).

\bibitem{dark_Werner} A. Werner, O.A. Egorov, and F. Lederer, Phys. Rev. B 85, 115315 (2012). 


\bibitem{darkS_Pinsker} F. Pinsker and H. Flayac, Phys. Rev. Lett. 112, 140405 (2014). 

\bibitem{SmirnovKivshar_PRB} L.A. Smirnov, D.A. Smirnova, E.A. Ostrovskaya, and Yu.S. Kivshar, Phys. Rev. B { \bf 89}, 235310 (2014). 



\bibitem{soliton_EgorovPRL} O. A. Egorov, D. V. Skryabin, A. V. Yulin, and F. Lederer, Phys. Rev. Lett. {\bf 102}, 153904 (2009).

\bibitem{solit_Egorov2010} O. A. Egorov, D. V. Skryabin, and F. Lederer, Phys. Rev. B {\bf 82}, 165326 (2010). 

\bibitem{solit_Egorov2012} O. A. Egorov, D. V. Skryabin, and F. Lederer, Phys. Rev. B, {\bf 84}, 165305 (2011). 

\bibitem{soliton_EgorovBook} O. A. Egorov, D. V. Skryabin, and F. Lederer, in {\it  Theory of Polariton Solitons}, edited by Z. Chen and R. Morandotti, Springer Series in Optical Sciences Vol. 170 (Springer,New York/Heidelberg, 2012), p. 171.

\bibitem{soliton_Sich} M. Sich, D. N. Krizhanovskii, M. S. Skolnick, A. V. Gorbach, R. Hartley, D. V. Skryabin, E. A. Cerda-Mendez, K. Biermann, R. Hey, and P. V. Santos, Nat. Photon. {\bf  6}, 50 (2012).

\bibitem{soliton_EgorovPRB} O. A. Egorov and F. Lederer, Phys. Rev. B {\bf 87}, 115315 (2013). 

\bibitem{soliton_SichPRL} M. Sich, F. Fras, J.K. Chana, M.S. Skolnick, D.N. Krizhanovskii, A.V. Gorbach, R. Hartley, D.V. Skryabin, S.S. Gavrilov, E.A. Cerda-Mendez, K. Biermann, R. Hey, and P.V. Santos, Phys. Rev. Lett. {\bf  112}, 046403 (2014) 



\bibitem{soliton_OstrovskayaPRA} E.A. Ostrovskaya, J. Abdullaev, A.S. Desyatnikov, M.D. Fraser, and Yu.S. Kivshar, Phys. Rev. A 86, 013636 (2012). 



\bibitem{vortex_Keeling} J. Keeling and N.G. Berloff, Phys. Rev. Lett. 100, 250401 (2008).

\bibitem{vortex_Marchetti} F. M. Marchetti, M. H. Szymanska, C. Tejedor, and D. M. Whittaker, Phys. Rev. Lett. 105, 063902 (2010).


\bibitem{vortex_Borgh} M.O. Borgh, G. Franchetti, J. Keeling, and N.G. Berloff, Phys. Rev. B 86, 035307 (2012).

\bibitem{vortex_Manni} F. Manni, T. C. H. Liew, K. G. Lagoudakis, C. Ouellet-Plamondon, R. Andre, V. Savona, and B. Deveaud, Phys. Rev. B 88, 201303(R) (2013).

\bibitem{vortex_Liew} T.C.H. Liew, Yuri G. Rubo, and A.V. Kavokin, Phys. Rev. Lett. 101, 187401 (2008). 



\bibitem{super_Tilley} D R Tilley and J Tilley, Superfluidity and Superconductivity(IOP publishing Ltd., Bristol, 1990) 




\bibitem{vortex_Pigeon} S. Pigeon, I. Carusotto, and C. Ciuti, Phys. Rev. B 83, 144513 (2011). 

\bibitem{vortex_Grosso} G. Grosso, G. Nardin, F. Morier-Genoud, Y. Leger, and B. Deveaud-Pledran, Phys. Rev. Lett. 107, 245301 (2011). 


\bibitem{solit_Grosso1} G. Grosso, G. Nardin, F. Morier-Genoud, Y. Leger, and B. Deveaud-Pledran, Phys. Rev. B 86, 020509(R) (2012).

\bibitem{vortex_Saito} Hiroki Saito, Tomohiko Aioi, and Tsuyoshi Kadokura, Phys. Rev. B 86, 014504 (2012) 


\bibitem{intrinsic}
J.A. Goldstone and E. Garmire, 
Phys. Rev. Lett.  {\bf 53}, 910 (1984). 




\bibitem{vortex_Voronova} N.S. Voronova, Yu.E. Lozovik, Phys. Rev. B {\bf 86}, 195305 (2012).



\bibitem{super_Tinkh} M. Tinkham, Introduction to Superconductivity, (McGraw-Hill Book Co., Singapore, 1996).


\end{thebibliography}
\end{document}